\documentclass[twoside,twocolumn,english,prl,superscriptaddress]{revtex4}
\usepackage[T1]{fontenc}
\usepackage{subfigure}
\usepackage{amsmath}
\usepackage{graphicx}

\makeatletter
\usepackage{bm}

\usepackage{babel}
\makeatother
\begin{document}

\title{Experimental Observation of the Inverse Spin Hall Effect at Room
Temperature}

\author{Baoli Liu}

\thanks{These authors contribute equally to this work.}

\affiliation{Beijing National Laboratory for Condensed Matter Physics, Institute
of Physics, Chinese Academy of Sciences, Beijing 100080, China}

\author{Junren Shi}

\thanks{These authors contribute equally to this work.}

\affiliation{Beijing National Laboratory for Condensed Matter Physics, Institute
of Physics, Chinese Academy of Sciences, Beijing 100080, China}

\affiliation{International Center for Quantum Structures, Chinese Academy of Sciences,
Beijing 10080, China}

\author{Wenxin Wang}

\affiliation{Beijing National Laboratory for Condensed Matter Physics, Institute
of Physics, Chinese Academy of Sciences, Beijing 100080, China}

\author{Hongming Zhao}

\affiliation{Beijing National Laboratory for Condensed Matter Physics, Institute
of Physics, Chinese Academy of Sciences, Beijing 100080, China}

\author{Dafang Li}

\affiliation{Beijing National Laboratory for Condensed Matter Physics, Institute
of Physics, Chinese Academy of Sciences, Beijing 100080, China}

\author{Shoucheng Zhang}

\affiliation{Department of Physics, Stanford University, Stanford, CA 94305, USA}

\author{Qikun Xue}

\affiliation{Beijing National Laboratory for Condensed Matter Physics, Institute
of Physics, Chinese Academy of Sciences, Beijing 100080, China}

\affiliation{International Center for Quantum Structures, Chinese Academy of Sciences,
Beijing 10080, China}

\author{Dongmin Chen}

\affiliation{Beijing National Laboratory for Condensed Matter Physics, Institute
of Physics, Chinese Academy of Sciences, Beijing 100080, China}

\affiliation{International Center for Quantum Structures, Chinese Academy of Sciences,
Beijing 10080, China}

\begin{abstract}
We observe the inverse spin Hall effect in a two-dimensional electron
gas confined in AlGaAs/InGaAs quantum wells. Specifically, we find
that an inhomogeneous spin density induced by the optical injection
gives rise an electric current transverse to both the spin polarization
and its gradient. The spin Hall conductivity can be inferred from
such a measurement through the Einstein relation and the Onsager relation,
and is found to have the order of magnitude of $0.5(e^{2}/h)$. The
observation is made at the room temperature and in samples with macroscopic
sizes, suggesting that the inverse spin Hall effect is a robust macroscopic
transport phenomenon. 
\end{abstract}
\maketitle
Electrically controlling spins has been a major theme for the recent
studies of the spin electronics, promising future information processing
devices integrating both the charge and spin operations~\cite{Prinz1998,Wolf2001,Zutic2004}.
As one of the most elementary transport processes involving both spins
and charges, the spin Hall effect (SHE), i.e., the transverse spin
transport induced by an electric field, is the focus of the recent
theoretical studies~\cite{Dyakonov1971,Hirsch1999,Murakami2003,Sinova2004}.
Experimentally, both the spin accumulation measurements~\cite{Kato2004,Wunderlich2005,Zhao2006}
and the electronic measurement~\cite{Valenzuela2006} have demonstrated
the convincing evidence for the existence of SHE. Electrical measurement
of the spin current has also been reported~\cite{Cui2006}.

In this Letter, we experimentally investigate the reciprocal effect
of SHE -- the inverse spin Hall effect (ISHE), in semiconductors.
ISHE complements the SHE, demonstrating the possibility of charge
transport driven by the spin degrees of freedom. Because ISHE is closely
related to SHE by the general principle of Onsager relation, which
dictates the symmetry of the transport coefficients for the SHE and
its reciprocal, our study also provides a supplemental evidence for
the existence of SHE. Moreover, it also provides an accessible way
to quantitatively determine the spin Hall conductivity of a sample.

It is instructive to have a general theoretical analysis for the relevant
phenomena. For an isotropic two-dimensional (2D) system with the time-reversal
symmetry (e.g., non-magnetic systems), the set of the equations describing
the macroscopic near-equilibrium (linear) charge and spin transports
can be written as,\begin{align}
\bm j_{\mathrm{tr}}^{c}= & -qD^{cc}\bm\nabla\left[n(\bm x)+\left(\frac{\partial n}{\partial\mu}\right)_{h}q\phi(\bm x)\right]\nonumber \\
 & -qD_{H}^{cs}\hat{\bm z}\times\bm\nabla\left[S_{z}(\bm x)-\left(\frac{\partial S_{z}}{\partial h_{z}}\right)_{\mu}h_{z}(\bm x)\right],\nonumber \\
\bm j_{\mathrm{tr}}^{s}= & -D^{ss}\bm\nabla\left[S_{z}(\bm x)-\left(\frac{\partial S_{z}}{\partial h_{z}}\right)_{\mu}h_{z}(\bm x)\right]\nonumber \\
 & -D_{H}^{sc}\hat{\bm z}\times\bm\nabla\left[n(\bm x)+\left(\frac{\partial n}{\partial\mu}\right)_{h}q\phi(\bm x)\right],\label{eq:transport eqs}\end{align}
where $\bm j_{\mathrm{tr}}^{c}$ ($\bm j_{\mathrm{tr}}^{s}$) denotes
the charge (spin) transport current, and $n(\bm x)$ ($S_{z}(\bm x)$)
is the local number (spin) density of the carriers, respectively.
$q$ is the charge of the carrier; $\hat{\bm z}$ is the unit vector
perpendicular to the 2D conducting plane, and the transport (diffusion)
coefficients are denoted by $D$ with the appropriate superscripts
and subscripts ($s$ denotes spin, $c$ denotes charge, $H$ denotes
the transverse Hall transport). The external fields are imposed by
the electric potential $\phi(\bm x)$ and the Zeeman field $h_{z}(\bm x)$.
For simplicity and without loss the generality, only the transport
induced by the $z$-component of the spin density is considered. The
validity of Eq.~(\ref{eq:transport eqs}) can be established through
the general considerations of symmetries, as well as the constraint
that the transport currents must vanish when the (spin) density distribution
is in the equilibrium against the external fields (i.e., $n(\bm x)=n_{0}-(\partial n/\partial\mu)_{h}q\phi(\bm x)$
and $S_{z}(\bm x)=S_{z}^{0}+(\partial S_{z}/\partial h_{z})_{\mu}h_{z}$).

Equation~(\ref{eq:transport eqs}) fully describes the near-equilibrium
charge and spin transport, for both the diffusion induced by the (spin)
density inhomogeneity and the transport induced by the external fields.
Within this framework, the inverse spin Hall effect is described by
$\bm j_{\mathrm{tr},\, H}^{c}=\sigma^{\mathrm{ISH}}\hat{\bm z}\times\bm F_{s}$
with the spin force being defined as the gradient of Zeeman field
$\bm F_{s}=g\mu_{B}\bm\nabla h_{z}(\bm x)$~\cite{Zhang2004,Shi2006}.
It can be related to the transverse charge diffusion induced by the
spin density inhomogeneity: \begin{equation}
\bm j_{\mathrm{tr,\, H}}^{c}=-qD_{H}^{cs}\hat{\bm z}\times\bm\nabla S_{z}(\bm x)\label{eq:Anomalous diffusion}\end{equation}
via an Einstein relation that relates the inverse spin Hall conductivity
$\sigma^{\mathrm{ISH}}$ with the corresponding diffusion constant
$D_{H}^{cs}$: \begin{equation}
\sigma^{\mathrm{ISH}}=(q/g\mu_{B})(\partial S_{z}/\partial h_{z})_{\mu}D_{H}^{cs}.\label{eq:Einstein relation}\end{equation}
Moreover, the Onsager relation dictates the duality between ISHE and
SHE~\cite{Shi2006}:\begin{equation}
\sigma^{\mathrm{SH}}=\sigma^{\mathrm{ISH}},\label{eq:Onsager Relation}\end{equation}
where $\sigma^{\mathrm{SH}}$ is the spin Hall conductivity ($\bm j_{\mathrm{tr},\, H}^{s}=\sigma^{\mathrm{SH}}\hat{\bm z}\times\bm E$).

Equations~(\ref{eq:Anomalous diffusion})--(\ref{eq:Onsager Relation})
outline the general theoretical ground for our experimental scheme:
if we can observe and experimentally establish the ISHE diffusion
process described by Eq.~(\ref{eq:Anomalous diffusion}), then the
ISHE and SHE can be related to such a measurement by the Einstein
relation (\ref{eq:Einstein relation}) and the Onsager relation (\ref{eq:Onsager Relation}).
We stress that the theoretical ground of our experimental scheme is
the direct result of the general principles of the non-equilibrium
thermodynamics, and should be valid for any macroscopic near-equilibrium
transport phenomena, regardless the microscopic details of the system.
On the other hand, building a microscopic transport theory consistent
to these general principles is a nontrivial issue yet to be fully
clarified~\cite{Shi2006}.

\begin{figure}
\subfigure[]{\includegraphics[width=0.36\columnwidth]{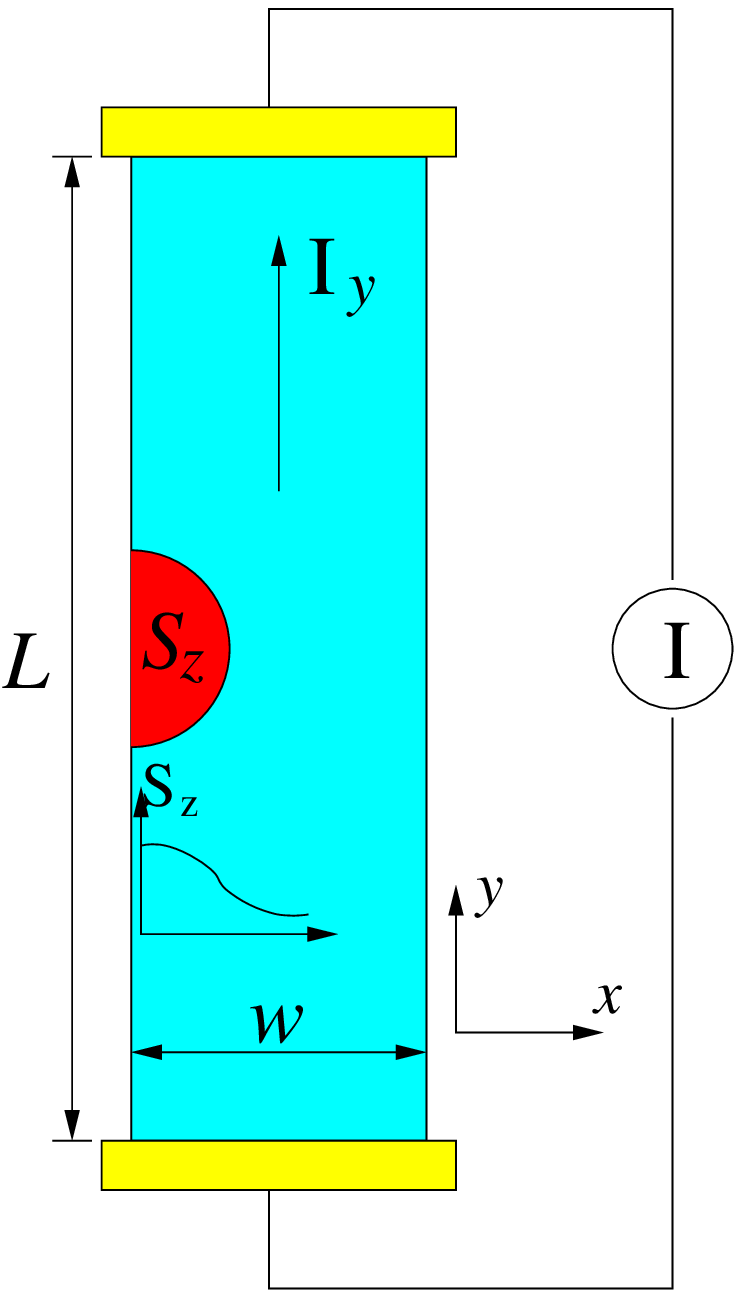}}\subfigure[]{\includegraphics[width=0.64\columnwidth]{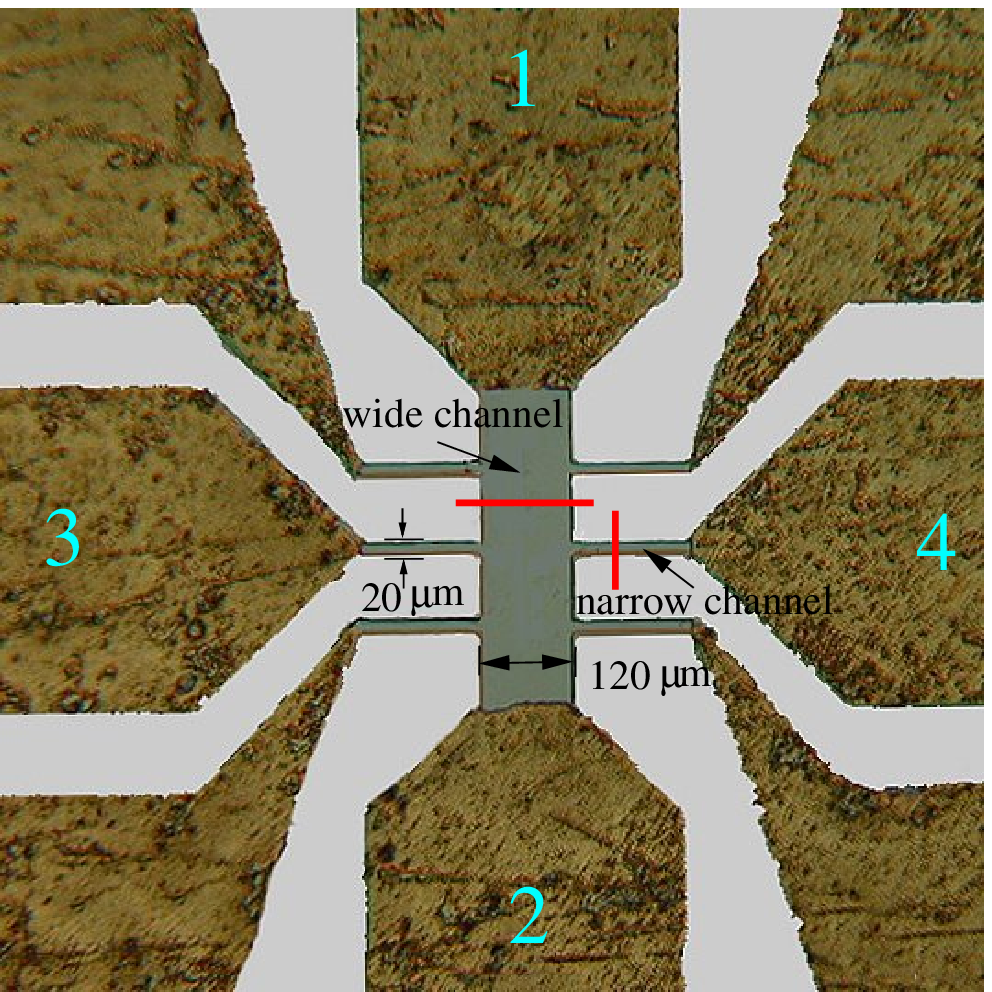}}

\caption{\label{fig:Experimental Scheme}(a) The schematics of the experimental
setup: an inhomogeneous spin density with Gaussian spatial profile
is created in a 2DEG channel using a circularly polarized light spot
with normal incidence. The transverse electric current induced by
ISHE is measured in the $y$-direction. (b) The lithography image
of the device: a multi-terminal Hall bar (indicated by the gray region)
with a $400\mu\mathrm{m}\times120\mu\mathrm{m}$ main channel and
six $150\mu\mathrm{m}\times20\mu\mathrm{m}$ arms, from which we obtain
one wide measurement channel by utilizing the main channel of the
Hall bar ($w=120\mu\mathrm{m}$, $L=400\mu\mathrm{m}$), and three
narrow measurement channels formed from three pairs of the arms ($w=20\mu\mathrm{m}$,
$L=300\mu\mathrm{m}$), as indicated in the image. The red lines indicate
the scanning paths of the light spot for obtaining the results presented
in Fig.~\ref{fig:CP}, and the electric current is measured through
the electrodes $1-2$ for the wide channel, and the electrodes $3-4$
for the narrow channel, respectively. }
\end{figure}

The experimental setup for observing the ISHE diffusion process is
shown schematically in Fig.~\ref{fig:Experimental Scheme}(a). A
circularly polarized laser light spot is employed to create an inhomogeneous
 spin density with Gaussian spatial profile~\cite{Meier1984}. An
overall gradient of spin density along $x$-direction is naturally
generated when the light spot is cutoff by the edges of the 2DEG channel,
resulting in asymmetric distribution of spin density. The electric
current due to the ISHE can then be measured along $y$-direction.
In such a setup, the transverse electric current is maximized when
the the light spot is centering at an edge of the channel, and reverses
its direction when centering at the opposite edge. The current vanishes
when the light spot is moved to the center of the channel, as the
spatial distribution of spin density is symmetric in this case. This
characteristic behavior of ISHE can be revealed by scanning the light
spot along $x$-direction to observe the dependence of the electric
current on the position of light spot. In theory, we find that total
transverse electric current is:\begin{equation}
I_{y}=-q\gamma D_{H}^{cs}\overline{\Delta S_{z}}\,,\label{eq:Iy}\end{equation}
where $\overline{\Delta S_{z}}=(1/L)\int_{0}^{L}\mathrm{d}y\Delta S_{z}(y)$
is the overall gradient of the spin density along $x$-direction,
averaged over the $y$-direction of the channel. $\Delta S_{z}(y)\equiv S_{z}(x=\mathrm{right\, edge},y)-S_{z}(x=\mathrm{left\, edge},y)$
is the difference of spin density at the right and the left edges
of the channel for given $y$, $L$ is the length of the sample, and
$\gamma=R_{c}/(R_{c}+R_{I})$ with $R_{c}$ and $R_{I}$ being the
resistance of the 2DEG channel and the measurement circuit, respectively~%
\footnote{Equation~(\ref{eq:Iy}) is a direct result of the more general formula
$I_{y}=-q\oint dy\rho_{c}(y)D_{H}^{cs}\Delta S_{z}(y)/\oint dy\rho_{c}(y)$,
where $\rho_{c}(y)$ is the resistivity, and the integration is  over
the closed loop of the circuit.%
}. 

The sample is grown by molecular beam epitaxy (MBE) on the semi-insulating
(001) GaAs substrate. 20 periods GaAs($2\mathrm{nm}$)/AlGaAs($6\mathrm{nm}$)
superlattices are first grown on the substrate, followed by $15\mathrm{nm}$
Al$_{0.24}$Ga$_{0.76}$As. 2DEG formed in a $13\mathrm{nm}$ In$_{0.19}$Ga$_{0.81}$As
quantum well, which is sandwiched between the barriers of $30\mathrm{nm}$
GaAs and $39\mathrm{nm}$ Al$_{0.24}$Ga$_{0.76}$As. Si $\delta$-doping
layers are buried in GaAs and Al$_{0.24}$Ga$_{0.76}$As barriers
with $10\mathrm{nm}$ and $4\mathrm{nm}$ spacer layers, respectively.
The standard Hall measurement gives the electron concentration $n=4.2\times10^{12}\mathrm{cm}^{-2}$
and the mobility $\mu=5000\mathrm{cm}^{2}/\mathrm{V}\cdot\mathrm{s}$
at the room temperature. Multi-terminal device of Hall bar geometry
is patterned using the standard photo-lithography and wet etching,
and the ohmic contacts are made with annealed AuGe/Ni. By utilizing
the main channel of the Hall bar and its six arms, we obtain one wide
channel and three narrow channels suitable for the proposed experiments,
as shown in Fig.~\ref{fig:Experimental Scheme}(b). All the measurements
are performed at the room temperature. Another sample with the same
geometry is also prepared on (110) GaAs substrate. Both samples yield
the similar results, and only the results for the (001) sample will
be presented in the following. 

\begin{figure}
\includegraphics[height=1\columnwidth,angle=270]{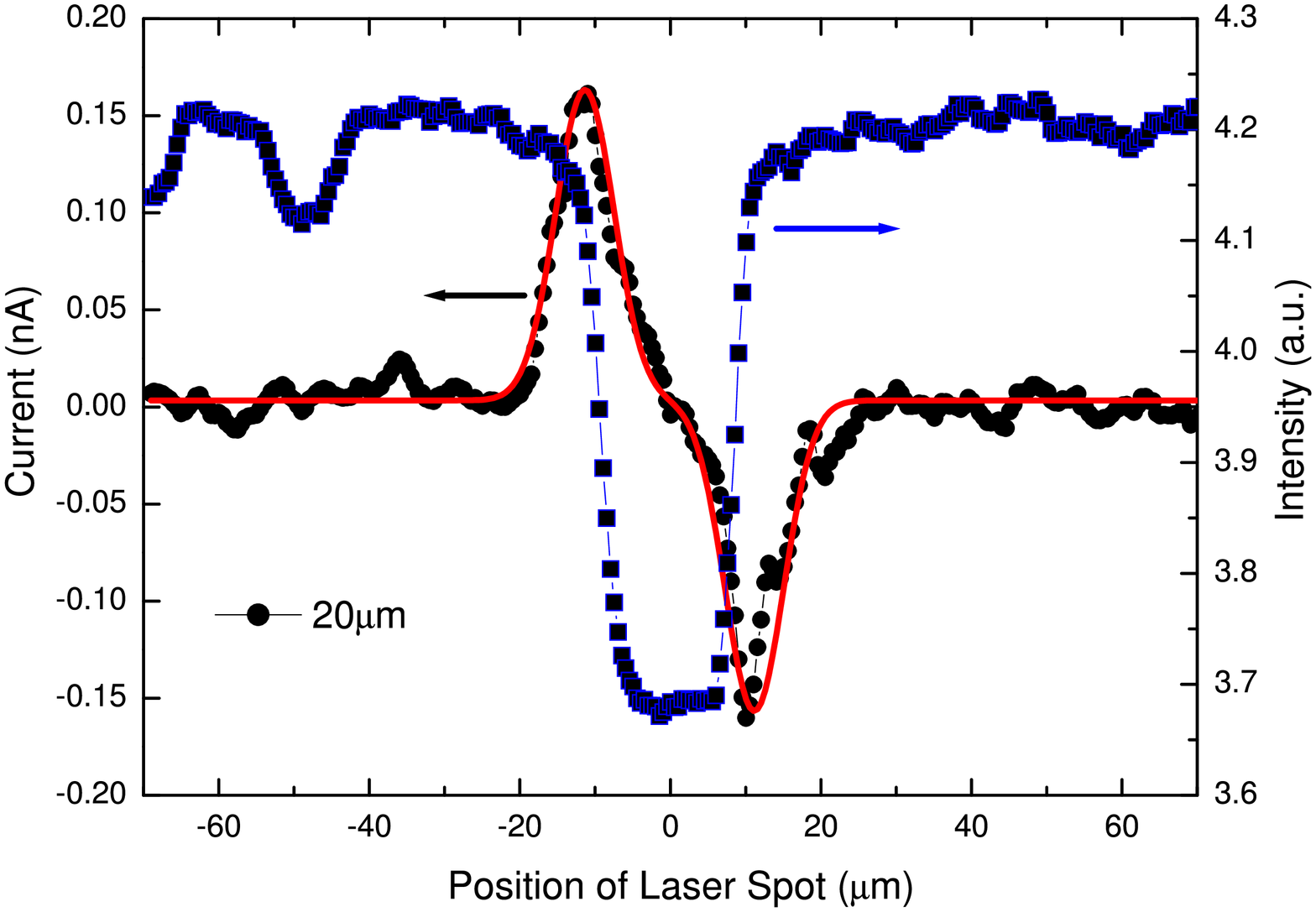}

\includegraphics[height=1\columnwidth,angle=270]{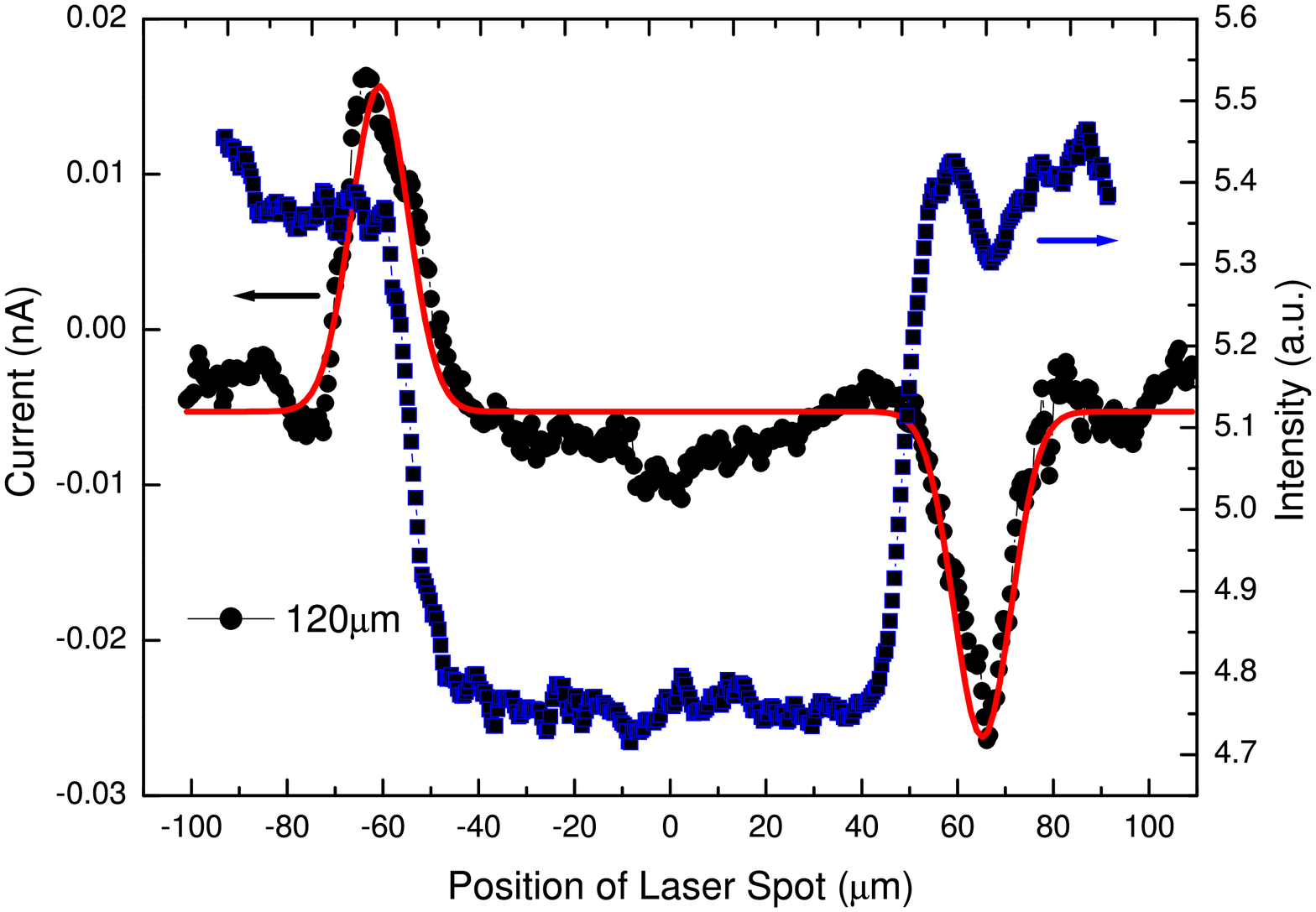}

\caption{\label{fig:CP}The light-spot-position dependence of the ISHE current
induced by a circularly polarized light, shown as black dots. The
blue-square curve shows the the intensity of reflected light: the
region occupied by the channel has the lower reflectivity due to the
additional absorption of light by the 2DEG. It thus provides a direct
sectional profile of the 2DEG channel along the scanning path, as
well as the precise determination of the position of the light spot
relative to the channel. Red curve shows the theoretical fitting to
the line-shape of electric current with the form described in the
text. The results for both a narrow channel (upper panel) and a wide
channel (lower panel) are shown.}
\end{figure}

In the experiment, a tunable continuous-wave Ti-Sapphire laser is
employed for the inter-band excitation with $20\mathrm{mW}$ average
pump power at the wavelength of $830\mathrm{nm}$. Its polarization
is modulated by a photo-elastic modulator (PEM), which yields a periodically
oscillating polarization between right- ($\sigma^{+}$) and left-
($\sigma^{-}$) circularly polarized light with a period of $50\mathrm{kHz}$
and a constant light intensity, resulting in a modulated spin density
and a constant carrier density~\cite{Koopmans1999}. The laser light
is focused to spot with full-width at half maximum (FWHM) of $8.0\mu\mathrm{m}$
through an objective ($\mathrm{N.A.}=0.38$), and is normal to the
sample. The spatial distribution of the $z$-component spin density
had been independently measured by the time and spatially resolved
Kerr rotation spectroscopy in such system, and was found to have the
Gaussian profile. The sample position is precisely controlled by a
motorized translation stage with $0.1\mu\mathrm{m}$ unidirectional
repeatability (Physik Instrumente, M-126.DG). By moving the sample,
the center of the light spot is scanned across the channel. The position
of the light spot on the sample is precisely determined through the
simultaneous measurement of the intensity of reflected light from
the sample surface, which provides a direct profiling of sectional
geometry of the channels along the scanning paths. The electric current
is measured directly by the lock-in amplifier, which effectively filters
out the electric current induced by effects not related to spins such
as the Dember effect and the photovoltaic effects, as the PEM only
modulates the spin density but keeps the carrier density constant. 

Figure~\ref{fig:CP} presents the main results of this study. For
both the wide and the narrow channels, the light-spot-position dependence
of the electric current shows the characteristic $N$-shape: it peaks
at the two edges with the opposite signs, and vanishes at the center
of the channel. The same behavior is observed in other scanning paths
and channels as well. The behavior is consistent with what the theory
expects from the ISHE diffusion process, Eq.~(\ref{eq:Anomalous diffusion}).
For a more quantitative analysis, we notice that the spin diffusion
length of the sample $\sim1\mu\mathrm{m}$ is much smaller than the
size of the light spot, thus the spin density $S_{z}(\bm x)$ excited
by the circularly polarized light can be considered to be proportional
to the local light intensity: $S_{z}(\bm x)\propto I(\bm x)$. The
intensity profile of a laser light spot is well described by the Gaussian
shape $I(\bm x)=I_{0}\exp\left[-(\bm x-\bm x_{0})^{2}/2\sigma^{2}\right]$,
where $\sigma$ is related to the FWHM of the laser spot by $\sigma=\mathrm{FWHM}/2\sqrt{2\ln2}\approx3.4\mu\mathrm{m}$.
The ISHE current has the form $I_{y}\propto\overline{\Delta S_{z}}\propto\exp\left[-x_{0}^{2}/2\sigma^{2}\right]\sinh\left(x_{0}w/2\sigma^{2}\right)$
, where $x_{0}$ is the center position of the light spot, and $w$
is the width of the channel. The form fits the line-shape of the experimental
data well. For the narrow channel, the fitting yields $w=22.5\pm0.2\mu\mathrm{m}$
and $\sigma=3.9\pm0.1\mu\mathrm{m}$, consistent with the independently
determined values $w=21\mu\mathrm{m}$ (determined from the intensity
of reflected light, blue square curves in Fig.~\ref{fig:CP}) and
$\sigma=3.4\mu\mathrm{m}$. For the wide channel, it yields $w=125.8\pm0.3\mu\mathrm{m}$
(vs. the independently determined value $w=123\mu\mathrm{m}$) and
$\sigma=6.2\pm0.2\mu\mathrm{m}$, and the somewhat over-estimated
$\sigma$ is due to the less sharp boundary of the edges (see the
blue square curve in the bottom panel of Fig.~\ref{fig:CP}). The
good correspondence between the experiment and the theory provides
a strong evidence for the existence of the ISHE. 

\begin{figure}
\includegraphics[height=1\columnwidth,angle=270]{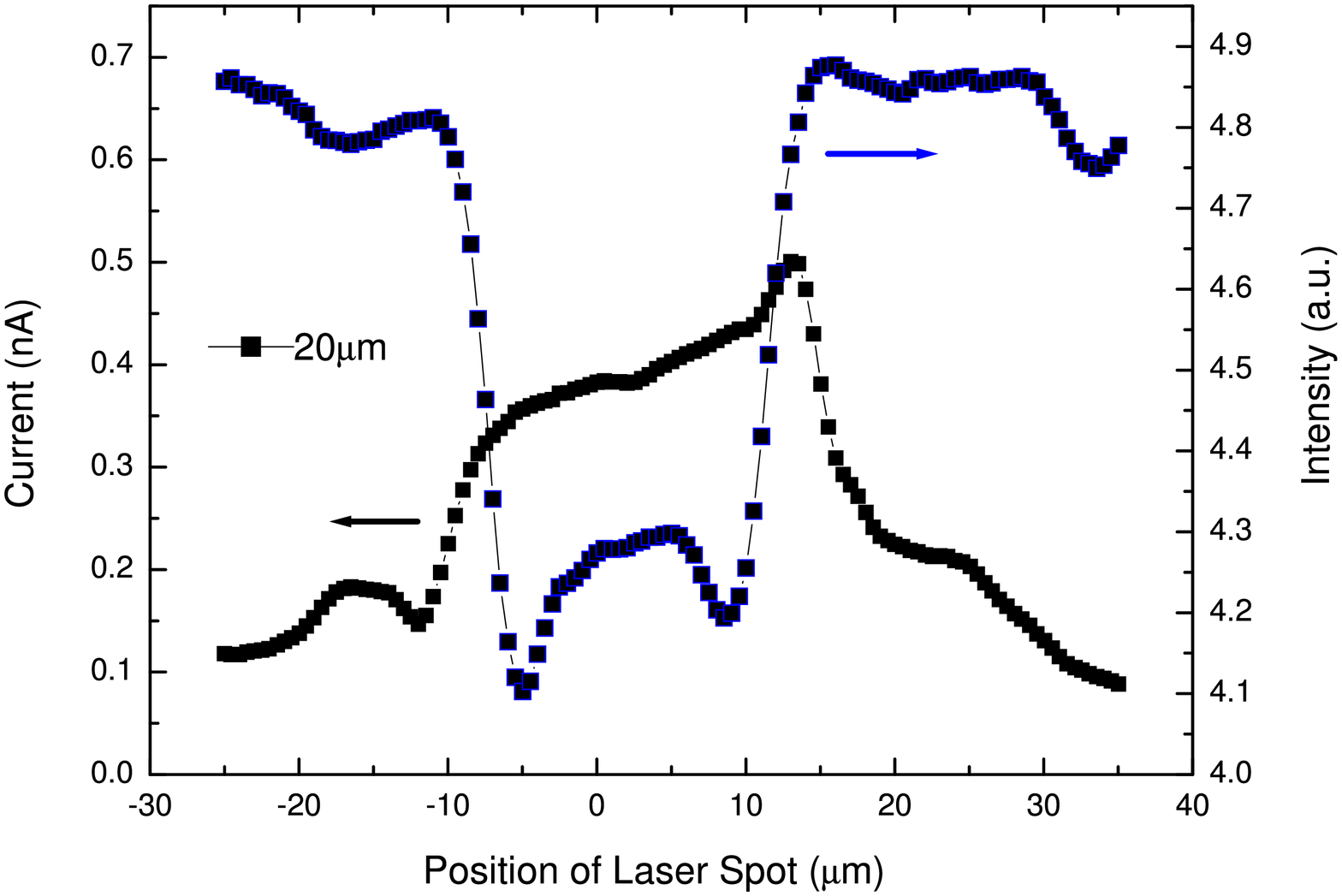}

\caption{\label{fig:LP}The electric current measured with the linearly polarized
light (black dots). The electric current measurement is locked-in
to the modulation of the light intensity (carrier density). The blue
squares indicate the intensity of reflected light.}
\end{figure}

While the observation can be naturally associated with the ISHE diffusion
process, it is necessary to rule out other possibilities that may
give rise the same behavior. We present the following points in order:
(i) the fact that the same behavior is observed in a number of different
channels and in a couple of different samples rules out the possibility
that it is caused by the asymmetry or inhomogeneity of the channels;
(ii) the fact that the electric current has the opposite signs at
the opposite edges of the channel indicates that the transport must
be driven by the gradient of the non-equilibrium density (either charge
or spin). This rules out the transport mechanisms that are driven
by the non-equilibrium density itself, for instance, the circular
photo-galvanic effect~\cite{Ganichev2001,Ganichev2000}, for which
the electric current is induced by the spin density, and flows along
a fixed direction resulted from the reduced symmetry of the sample;
(iii) the circularly polarized light employed in the measurement excites
the charge density as well as the spin density. This rises the possibility
that the phenomenon is induced by the gradient of the charge density
instead of the spin density, i.e., the modulation of the circular
polarization may induce a modulation of the charge density, which
in turn generates an electric signal picked up by the lock-in amplifier.
While this is highly improbable in theory because the sample has the
time-reversal symmetry which prohibits the charge Hall transport,
the possibility can be explicitly ruled out by a controlled experiment:
in Fig.~\ref{fig:LP}, the linearly polarized light is employed to
excite a pure charge density. It is evident that the incriminating
$N$-shape characteristics seen in the circularly polarized light
measurement (Fig.~\ref{fig:CP}) can no longer be observed. All summarized,
we conclude that the electric current measured in Fig.~\ref{fig:CP}
can only be induced by the gradient of spin density, as predicted
by Eq.~(\ref{eq:Anomalous diffusion}).

Finally, we can quantitatively determine the (inverse) spin Hall conductivity
based on such a measurement. For $20\mathrm{mW}$ light power, the
total injection rate (I.R.) of spins is estimated to be $3.9\times10^{14}\,\mathrm{spins}/\mathrm{s}$~%
\footnote{$\mathrm{I.R.}=(1/2)P/(h\nu)(T\alpha d)$, where $h\nu\approx1.5\mathrm{eV}$,
$T\approx0.6$ is the transmission rate of the light path, $\alpha d=4\pi kd/\lambda\approx0.016$
is the absorption rate of the 2DEG ($k=0.08$, $d=13\mathrm{nm}$
is the thickness of the quantum well) . The extra factor of $1/2$
is due to the fact that both the heavy hole and the light hole states
are excited .%
}. The total number of the injected spins $S_{\mathrm{tot}}=\mathrm{I.R.}\times\tau_{s}\approx3.9\times10^{4}\,\mathrm{spins}$
with a spin relaxation time $\tau_{s}\approx100\mathrm{ps}$. By using
Eq.~(\ref{eq:Iy}), the magnitude of the ISHE diffusion constant
can be estimated by $D_{H}^{cs}\approx\sqrt{2\pi}I_{y}^{\mathrm{max}}\sigma L/(e\gamma S_{\mathrm{tot}})$
(assuming $w\gg\sigma$), and yields values $D_{H}^{cs}(20\mu\mathrm{m})\approx0.87\mathrm{cm}^{2}/\mathrm{s}$
and $D_{H}^{cs}(120\mu\mathrm{m})\approx0.4\mathrm{cm}^{2}/\mathrm{s}$,
respectively~%
\footnote{The parameters used: for $20\mu\mathrm{m}$ channel: $R_{c}=6.2\mathrm{k}\Omega$,
$I_{y}^{\mathrm{max}}=0.16\mathrm{nA}$, $L=300\mu\mathrm{m}$; for
$120\mu\mathrm{m}$ channel: $R_{c}=1.1\mathrm{k}\Omega$, $I_{y}^{\mathrm{max}}=0.021\mathrm{nA}$,
$L=400\mu\mathrm{m}$. The measurement circuit has an inner resistance
$R_{I}=1\mathrm{k}\Omega$.%
}. By applying the Einstein relation Eq.~(\ref{eq:Einstein relation}),
$\sigma^{\mathrm{SH}}\approx(ne/kT)D_{H}^{cs}$ ($\partial S_{z}/\partial h_{z}\approx g\mu_{B}n/kT$),
we get the spin Hall conductivity (in the unit of charge conductivity):
$\sigma^{\mathrm{SH}}(20\mu\mathrm{m})\approx0.59(e^{2}/h)$ and $\sigma^{\mathrm{SH}}(120\mu\mathrm{m})\approx0.27(e^{2}/h)$,
respectively. The spin Hall conductivities determined from the two
different channels yield consistent values (within a factor of $2$),
indicating the self-consistency of our method of analysis. We note
that all the parameters involved in the estimate can in principles
be precisely determined. Our experimental scheme can thus provide
an accessible way to measure the spin Hall conductivity. 

In summary, we have observed the ISHE diffusion process, experimentally
establishing one of the elementary processes of the spin-charge transports.
Because ISHE is closely related to SHE by the general principle of
Onsager relations, the observation provides a supplemental evidence
for the existence of the spin Hall effect. Moreover, our experimental
scheme provides an accessible way to quantitatively determine the
spin Hall conductivity of the system, implying the wider application
of the technique in the future studies of the spin-charge transports.
Finally, the fact that the measurement is carried out at the room
temperature and in samples with macroscopic sizes suggests that ISHE
is a robust macroscopic transport phenomenon -- a universal property
that presents in various systems.

We gracefully acknowledge the useful discussions with Qian Niu and
Yugui Yao, and the assistance of Changzhi Gu for preparing the devices.
This work was supported by the Knowledge Innovation Project of the
Chinese Academy of Sciences, and NSFC-10534030. J.R.S. was supported
by the {}``BaiRen'' program of the Chinese Academy of Sciences.

\bibliographystyle{/usr/local/share/texmf/tex/latex/Science}
\bibliography{/home/shi/Papers/Papers}

\end{document}